\documentclass[prb,floatfix,reprint]{revtex4-2}

\usepackage{amsmath}
\usepackage{amssymb}
\usepackage{graphicx}
\usepackage{hyperref}
\usepackage{color}
\usepackage{bm}

\begin{document}

\title{Photon counting beyond the rotating-wave approximation}

\author{Steven Kim and Fabian Hassler}
\affiliation{Institute for Quantum Information, RWTH Aachen University, 52056 Aachen, Germany}
\date{June 2026}
\begin{abstract}
Open quantum systems are often described by a Lindblad master equation, which relies on a set of approximations, most importantly the rotating-wave approximation which is only valid for weak damping.
In the Lindblad setting, dissipative processes are described through jump operators, distinguishing between absorption and emission of photons. 
This enables the simple identification of emitted photons which provides a straightforward way to obtain the radiation statistics.
Outside the rotating-wave limit, the Lindblad approach does not work.  Open quantum systems can then be described by, \emph{e.g.}, the quantum Langevin equation.
However, in this framework the number of emitted photons is not easily accessible.
In this work, we point out how to obtain the photon counting statistics from a quantum Langevin equation and provide an expression for the photon current operator, for arbitrary systems coupled to linear environments.
As an example, we employ the  method to study the radiation statistics of a damped harmonic oscillator at finite temperature beyond the rotating-wave approximation.
We show that even outside the rotating-wave limit, the most important contribution to the radiation statistics can be captured by an effective Lindblad equation, thus extending the range of possible applications of the Lindblad framework.

\end{abstract}
\maketitle

\section{Introduction}
Open quantum systems interact with their environment, \textit{e.g.}, via the absorption or emission of photons.
Often, such systems are described by a Lindblad master equation which provides a simple local in time description of the dynamics of the system \cite{campaioli}.
For the Lindblad equation to be valid, one needs that bath-system correlations can be neglected and that correlations in the bath decay much quicker than correlations in the system, also known as the Born-Markov approximation.
Further, weak damping is necessary which enables a rotating-wave approximation (RWA) that reduces the complexity of the problem only to the resonance frequency of the system.

For systems such as optical cavities, this approximation is essentially always valid due to the large frequencies involved \cite{haroche}.
However, for systems that operate at lower frequencies, the resonance frequency can become of the order of the dissipation rate. 
This can be the case for superconducting circuits that operate in the microwave regime \cite{westig:17}. 
Additionally when the number of emitted photons is small, a strong-coupling provides an advantage of an increased signal.
In such situations, the RWA is not applicable anymore and the description by a Lindblad equation is not valid.

More generally, open quantum systems can be described by a quantum Langevin equation (QLE), earning its name due to its resemblance to the classical stochastic Langevin equation \cite{caldeira, ford:88}.
The advantage of the QLE compared to the Lindblad equation is that it does not rely on approximations such as  Markovian environments or a RWA.
However, all dissipative processes are collected in a damping kernel, making the distinction between different dissipative processes hard.
This stands in contrast to the Lindblad setting where each process is uniquely described by a jump operator.
In particular, this enables an easy identification of the emission of photons which can be used to obtain the radiation statistics \cite{davies:81, kim:23, kansanen:25}. 
Thus, the question arises how to obtain radiation statistics from the QLE.

In the past, QLEs have  been used to study radiation statistics in the context of input-output theory \cite{gardiner:io, ciuti, roy:18}.
However, the results rely on a RWA or Markovian environments. 
Beyond these limitations, the QLE has been employed to extensively study the damped harmonic oscillator where a diverging velocity correlation was predicted \cite{devoret:96, weiss}.
However, it is important to note that the studies  studied the state inside the system and have not been applied in the context of radiation statistics which concerns the emission into the environment.

Furthermore, the QLE naturally adopts non-Markovian dynamics both through damping kernels with a finite correlation time and time-correlated colored noise.
This enables the study of non-Markovian dynamics which is recently the focus of many studies \cite{nori, vacchini, alonso, lednev:24, ankerhold:26}.
An alternative approach to the QLE is based on, \emph{e.g.}, auxillary bosonic modes that collectively reproduce the correct damping kernels, see Ref.~\cite{lednev:24}. 

In this paper, we study how to obtain the photon counting statistics of an arbitrary system beyond the RWA using the QLE.
In Sec.~\ref{sec:QLE}, we provide a quick derivation of the QLE and explain relevant concepts.
Next, we provide an expression of a photon current operator that enables the study of radiation statistics.
In Sec.~\ref{sec:statistics}, we apply our method to the damped harmonic oscillator and study the radiation statistics beyond the RWA.
Further, we also show how to derive an effective Lindblad equation providing good approximation also outside the rotating-wave limit.

\section{Quantum Langevin equation}\label{sec:QLE}
\begin{figure}[tb]
	\centering
	\includegraphics[scale=0.9]{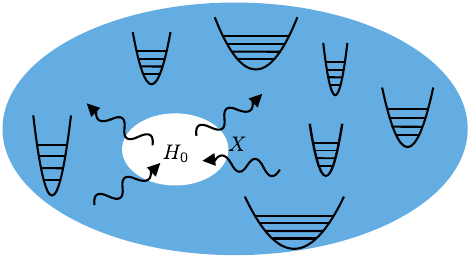}
	\caption{%
	Sketch of the Caldeira-Leggett model. A system with Hamiltonian $H_0$ is coupled to an environment that consists of a bath of harmonic oscillators.
	The  coupling of the system operator $X$ to the bath modes  allows for an exchange of energy.
	When tracing out the environment, the dynamics of the system can be described by a QLE, see Eq.~(\ref{eq:QLE}).
}\label{fig:caldeira}
\end{figure}
Setups where the photon counting statistics can be of interest range from, \emph{e.g.}, cavity or circuit quantum electrodynamics (cQED) setups to optomechanics and nanomechanical oscillators \cite{carmichael, blais, aspelmeyer, nunnenkamp, clerk, schwab, cleland}.
We use the term `photon' to label the emission of quantized energy which appear in different forms in these platforms, \emph{e.g.}, in nanomechanical oscillators, they correspond to phonons.
The setting is that an arbitrary system interacts with a (typically thermal) environment, which leads to dissipation, \emph{i.e.}, an exchange of energy between system and environment.
Additionally, the system can be driven where the driving mechanism can take many forms, ranging from external pumping \cite{portugal:23}, to normal and superconducting tunnel junctions \cite{portier:10, menard:22}, or also spin-boson interactions \cite{ritsch}.

As the system emits energy into the environment, the environment also acts as a detector.
The radiation can be collected and quantified, revealing insights into, \emph{e.g.}, entanglement and strong- or anti-correlations \cite{peugeot:21, kim:24, wallraff:11, loudon}.
For this purpose, it is necessary to find an appropriate description of the environment and the system interacting with it.

The environment can be described as a bath of harmonic oscillators with mass $m_n$ which are all linearly coupled with strength $k_n$ to the system, also known as the Caldeira-Leggett model \cite{leggett}, see Fig.~\ref{fig:caldeira}.
The total Hamiltonian of system and environment combined is given by 
\begin{equation}\label{eq:CL-model}
  H = H_0 + \sum_n \left[\frac{p_n^2}{2m_n} + \frac{k_n}{2}(q_n-X)^2\right]
\end{equation}
where $p_n$ and $q_n$ are momentum and position operators of the bath oscillators with the canonical commutator $[q_i, p_j]= i\hbar\delta_{ij}$.
Here, the coupling operator $X$ is a position-like operator of the system with Hamiltonian $H_0$. Most importantly, $X$ should not be diagonal in the energy basis in order to enable an exchange of energy.
For example, for a particle of mass $m$ in a potential $V(x)$ with Hamiltonian $H_0 = p^2/2m + V(x)$, $[x,p]=i\hbar$, the coupling operator typically is the position $X=x$.
For a two-level system with the Hamiltonian $H_0 = \hbar\Omega \sigma_z/2$, the coupling operator is given by $X = \sqrt{\hbar/\Omega}\,\sigma_x$, where $\sigma_i$ are Pauli matrices.

If the system-bath coupling is assumed to be small, an effective description of the system in the form of  a Lindblad or Bloch-Redfield equation can be obtained \cite{breuer}. 
Recently, it has been shown that, even in the strong coupling regime, it is possible to derive a Lindblad equation \cite{lednev:24}.
This, however, requires adding additional bosonic auxiliary modes and thus increases the complexity.
However, every formulation in Lindblad form relies on a RWA.

Here, we follow an alternative route to obtain an effective description of the system, starting from the Hamiltonian~(\ref{eq:CL-model}).
In the following, we briefly present the main idea of the method, details can be found in Ref.~\cite{gardinerzoller}.
Note that the Hamiltonian describes a linear environment.
Thus, it is possible to solve the Heisenberg equations of motion, $\dot p_n = i[H, p_n]/\hbar$ and equivalently for $\dot q_n$, exactly.
We insert the solutions into the Heisenberg equation of an arbitrary, time-independent system operator $Y$ to obtain
\begin{equation}\label{eq:QLE}
\dot{Y} = \frac{i}{\hbar}[H_0, Y]-\frac{i}{2\hbar}\Bigl\{[X,Y], \xi(t)-\int_{-\infty}^t f(t-t')\dot X(t') dt' \Bigr\}.
\end{equation}
This equation is dubbed QLE due to its similarity to  a classical Langevin equation.
Note that this expression can be straightforwardly extended to include multiple systems and baths.

The first term describes the coherent time evolution for a closed system with Hamiltonian $H_0$.
The additional term stems from the interaction with the environment.
Here, $\xi(t)$ represents the noise and purely depends on the environment.
The other contribution connects to $\dot{X}$ and, thus, describes the damping.
The nature of the damping is encoded in the damping kernel $f(t)$.
While the damping kernel can be given in terms of the system-bath coupling constants $k_n$, it is not necessary to specify the microscopic details of the coupling.
It suffices to provide the damping kernel $f(t)$ as it relates to the noise operator $\xi(t)$ via fluctuation-dissipation relations.

The fluctuation-dissipation relations are revealed by the commutation relations of the noise operator $\xi(t)$, which for a thermal environment at temperature $T$ read
\begin{equation}\label{eq:noise_symm}
\tfrac12 \langle \{\xi_\omega, \xi_{\omega'}\} \rangle = 2\pi \hbar\omega \mathrm{Re}(f_\omega) (2 n_\omega + 1)\delta(\omega+\omega')
\end{equation}
and
\begin{equation}
[\xi(t),\xi(t')] = i \hbar \frac{d}{dt} f(t-t').
\end{equation}
Here, we have introduced the Fourier transform $f_\omega = \int dt f(t) e^{i\omega t}$ and the Bose-Einstein distribution $n_\omega = [\exp(\hbar\omega/k_B T) -1]^{-1}$ with $k_B$ the Boltzmann constant.
Note that $\xi(t)$, in general, represents correlated noise.
As such, even for local-in-time damping kernels $f(t) \sim \delta(t-0^+)$ \cite{note_damping}, the time evolution shows memory effects.

It is important to note that the QLE describes the whole dynamics of the system. 
In particular the contribution that leads to damping, $\int \! f(t-t') \dot X(t') dt'$, does not separate different dissipative processes such as emission or absorption of photons.
This stands in contrast to the Lindblad equation where different dissipative processes are described by different jump operators.
Thus, the question arises how to extract observables such as the number of emitted photons which requires a description by ladder operators.
This will be treated in the following section.

\section{Radiation statistics}\label{sec:photon}
For very weakly damped systems, the spectral linewidth is narrow and the system emits photons mainly at its characteristic frequency.
For larger damping however, the linewidth increases and thus the systems emits photons in a broad range of frequencies.
To be able to study the radiation statistics of a system, we require a photon current operator $I(t)$.
This goal can be achieved by finding a frequency and time resolved power output $P_\omega(t)$ which can be related to a frequency resolved photon current via $I_\omega(t) = P_\omega(t)/\hbar\omega$.
The total photon current can be then obtained by integrating over all frequencies $I(t) = \int\! I_\omega(t)d\omega/2\pi$.

To obtain the emitted power, we need to compute $\dot H_0$, \emph{i.e.}, the rate of change of energy of the system.
From Eq.~(\ref{eq:QLE}), we obtain
\begin{equation}
\dot H_0 = \frac12 \Bigl\{ \dot X(t), \xi(t) - \int_{-\infty}^t f(t-t')\dot X (t')dt' \Bigr\},
\end{equation}
where $\dot X = i[H_0, X]/\hbar$ has been employed.
The contribution of the noise $\xi(t)$ describes incoming energy fluctuations due to the environment.
The second term is due to dissipation, \emph{i.e.}, photon emission, and thus provides us with the power 
\begin{equation}\label{eq:power}
P(t) = \frac12 \Bigl\{ \dot X(t), \int_{-\infty}^t f(t-t')\dot X (t')dt' \Bigr\}.
\end{equation}
By moving to frequency space, we can relate the power to a frequency resolved expression in the frequency window $[\omega, \omega+d\omega]$ via $dP(t) = P_\omega(t)d\omega/2\pi$.
As described above, this also provides the frequency resolved photon current $I_\omega(t) = P_\omega(t)/\hbar\omega$ which can be used to obtain the emitted photon current  $I(t) = \int\! I_\omega(t)d\omega/2\pi$.
We arrive at the central result
\begin{equation}\label{eq:photon_current}
I(t) = \frac1\hbar\int\!\!\!\!\int \frac{d\omega d\nu}{(2\pi)^2} \operatorname{Re}(f_\omega) A_{\omega+\nu/2}^\dag A_{\omega-\nu/2}e^{i\nu t},
\end{equation}
where $A_\omega = \Theta(\omega)\sqrt{|\omega|}X_\omega$ with the step function $\Theta(\omega)$, $A_\omega^\dag = A_{-\omega}$, and we have assumed that $\omega t\gg 1$. 
Note that the photon current can also be modulated with a detection efficiency $0\leq\eta_\omega\leq1$.

In the following, we are going to explain the different parts of this expression.
First, $\operatorname{Re}(f_\omega)$ provides a frequency dependent emission rate.
For example, in the case of Ohmic damping $f(t) \propto \gamma \delta(t-0^+)$ we have that $\operatorname{Re}(f_\omega)\propto \gamma$.
Thus, photons of all frequencies are equally emitted at the rate $\gamma$. 

The annihilation operator  $A_\omega = \Theta(\omega)\sqrt{|\omega|}X_\omega$ describes the emission of photons at the (continuous) frequency $\omega$. 
The projection onto positive frequencies is crucial, to select the emission processes. 
In the context of input-output theory, $A_\omega$ relates to the output field while $\xi(t)$ describes the incoming field.
The negative frequencies describes absorption due to the relation $A_{-\omega} = A_\omega^\dag$ \cite{glauber, clerk, lesovik:97}.

Note that in this form, $A_\omega$ is not dimensionless and is only proportional to the photon ladder operators $a_\omega$ that are conventionally used.
The exact relation between $A_\omega$ and $a_\omega$ depends on the system.
For the examples mentioned above, $H_0 = p^2/2m + V(x)$ and $H_0 = \hbar \Omega \sigma_z/2$, normalized ladder operators can be obtained by $a_\omega = \sqrt{m/2\hbar}A_\omega$ and $a_\omega = (\sqrt{\hbar}/\Omega) A_\omega$ respectively.
Then, they describe photons \cite{roy:18}.
However, normalized correlation functions are independent of this prefactor, see below.

`Energy-time' fluctuations lead to a frequency  uncertainty described by the frequency difference $\nu$, which become important when intensity correlations are studied \cite{hassler:15}.
In this respect, the expression for the photon current $I(t)$ in Eq.~(\ref{eq:photon_current}) is akin to a Wigner transform \cite{nazarov:03, hassler:15}.

Equipped with the photon current operator $I(t)$ expressed via the (unnormalized) annihilation operators $A_\omega$, we can study  the photon statistics of the radiation emitted into the environment.
In particular, the cumulants of the number of emitted photons
\begin{equation}\label{eq:number}
N(\tau) = \int_0^\tau  I(t) dt
\end{equation}
during the detection time $\tau$ are of interest.
The average number of emitted photons connects to the average emitted photon current via $\langle N \rangle = \langle I \rangle \tau$ and grows linearly with the detection time in a stationary situation.
The higher-order cumulants can be obtained from time-dependent coherence functions.
For example, the Fano factor $F = \langle \! \langle N^2\rangle \! \rangle/ \langle N \rangle$, that measures the number of correlated photons, can be obtained from the second-order coherence 
\begin{equation}\label{eq:g2}
g^{(2)}(\tau)=\frac{\langle :\!I(\tau) I(0)\!: \rangle }{\langle I \rangle^2}
\end{equation}
via the relation $F= 1+ \langle I \rangle \int d\tau [g^{(2)}(\tau)-1]$ \cite{emary}.
Similarly, higher-order cumulants are connected to higher-order correlators.
The colons in Eq.~(\ref{eq:g2}) denote the conventional normal ordering procedure: all $A^\dag$ to the left of all $A$ where the $A$ are time-ordered and $A^\dag$ are anti-time-ordered.
The normal ordering, in particular, cancels potential vacuum fluctuations.

As an example of the formalism, we are going to study the radiation statistics of a damped harmonic oscillator beyond the RWA in the following section.

\section{Damped Harmonic Oscillator}\label{sec:statistics}
\begin{figure}[tb]
	\centering
	\includegraphics{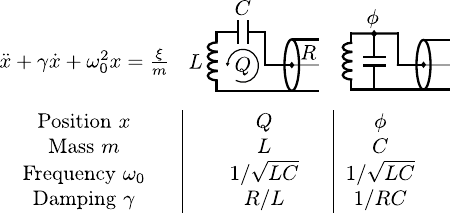}
	\caption{%
	Two setups from cQED which are described by the QLE of a damped harmonic oscillator. The oscillators consist of an inductance $L$, capacitance $C$, and Ohmic resistor $R$ depicted as a transmission line. For the in-series circuit (left), the loop charge $Q$ takes the place of the position, while in the parallel case (right), this role is played by the node flux $\phi$.
}\label{fig:circuit}
\end{figure}
In this section, we study the radiation statistics of the damped harmonic oscillator beyond the RWA.
The system is described by the Hamiltonian $H_0 = p^2/2m + m\omega_0^2 x^2/2$ with the mass $m$, characteristic frequency $\omega_0$, momentum $p$, and position $x$ with the canonical commutator $[x,p]=i\hbar$.
The system couples to the environment via $X=x$ and the damping kernel is Ohmic with $f(t) = m\gamma \delta(t-0^+)$.
Examples from cQED that realize this system are depicted in Fig.~\ref{fig:circuit}.

With Eq.~(\ref{eq:QLE}), the QLE of the damped harmonic oscillator is given by
\begin{equation}\label{eq:x_qle}
m(\ddot x +\gamma \dot x + \omega_0^2 x) = \xi(t)
\end{equation}
where $p=m\dot x$ has been used.
The solution of this equation of motion can be obtained by employing a Fourier transform which leads to $x_\omega = \xi_\omega/(-\omega^2-i\gamma\omega+\omega_0^2)m$.
With this, we also obtain the photon emission operator $A_\omega = \Theta(\omega)\sqrt{|\omega|}x_\omega$.
For the following, we use the normalized operator $a_\omega = \sqrt{m/2\hbar} A_\omega$, see above.

As described in the previous section, coherence functions are the central topic when studying radiation statistics.
In the case of the harmonic oscillator, more generally a chaotic light source \cite{loudon}, it can be shown that all coherence functions can be related to the first-order coherence
\begin{equation}
g^{(1)}(\tau)=\frac{\langle a^\dag(0)a(\tau) \rangle }{\langle a^\dag a \rangle}
\end{equation}
with $g^{(1)}(0)=1$.
In particular, the second-order coherence fulfills the Siegert relation $g^{(2)}(\tau) = 1 + |g^{(1)}(\tau)|^2$.
The first-order coherence can be measured using a homodyne detection scheme, see \emph{e.g.}, Ref.~\cite{wallraff:10}.

Because of this, our focus will be the first-order coherence.
More specifically, we define the correlator 
\begin{equation}
G^{(1)}(\tau) = \langle a^\dag(0) a(\tau)\rangle 
\end{equation}
which also provides the average photon current via $\langle I \rangle = \gamma G^{(1)}(0)$.
Using the solution of Eq.~(\ref{eq:x_qle}), see above, with the normalized ladder operator $a_\omega = \Theta(\omega)\sqrt{m|\omega|/2\hbar}\,x_\omega$, we arrive at
\begin{equation}\label{eq:G1}
G^{(1)}(\tau) = \int_0^\infty \frac{d\omega}{2\pi} \frac{4 \gamma \omega^2 n_\omega}{(\omega^2-\omega_0^2)^2+\gamma^2\omega^2}e^{-i\omega \tau}.
\end{equation}

\subsection{Comparison to previous studies}
In previous studies of the QLE of the damped harmonic oscillator, see \emph{e.g.} Ref.~\cite{devoret:96, weiss}, system correlators like $\langle\!\langle x^2 \rangle\!\rangle$ and $\langle\!\langle p^2 \rangle\!\rangle$ have been calculated.
These correlators are not well-defined as, \emph{e.g.}, $\langle\!\langle p^2 \rangle\!\rangle$ has a logarithmic divergence at large frequencies. Moreover, they are not obviously related to measurable quantities.
In particular, by defining the conventional annihilation operator  $b = \sqrt{m\omega_0/2\hbar}(x+ i p/m\omega_0)$  for the photons in the cavity, a diverging number of photons is obtained.
In contrast, Eq.~(\ref{eq:G1})  converges without additional cutoff. This leads to a finite   photon current $\langle I\rangle =\gamma G^{(1)}(0)$ that is emitted into the environment.

The divergence arises due to the missing separation of positive and negative frequencies. 
The QLE for $b$ is given by
\begin{equation}
\dot b = -i\omega_0 b -\frac{\gamma}{2}(b-b^\dag) + \frac{i}{\sqrt{2\hbar m \omega_0}}\xi.
\end{equation}
In this form, it is evident that $b$ couples to $b^\dag$ which describes the gain of photons and thus leads to an artificial pumping of the mode \cite{lednev:24}.
This in turn leads to a diverging number of `photons' $\langle b^\dag b \rangle$, which also explains the divergence of $\langle\!\langle p^2\rangle\!\rangle$ \cite{devoret:96, weiss}.

Note that in the rotating-wave limit $\gamma \ll \omega_0$, the term proportional to $b^\dag$ is neglected as it is counter-rotating contribution. Because of this, no divergence arises in the conventional Lindblad description.
In the following, we will show that our formalism recovers the Lindblad result in the rotating-wave limit $\gamma \ll \omega_0$ where the system only emits at the resonance frequency $\omega_0$. 
Moreover, it remains finite for stronger coupling where the photon emission is distributed in a broader frequency range.

\subsection{Pole structure}
\begin{figure}[tb]
	\centering
	\includegraphics{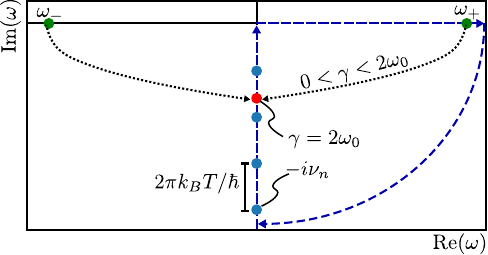}
	\caption{%
	  Pole structure of the integrand in Eq.~(\ref{eq:G1}) in the complex $\omega$-plane. The poles $\omega_\pm$ (green) start on the real axis in the rotating-wave limit $\gamma \to 0^+$ and  approach the imaginary axis close to the critical damping. At the critical point $\gamma=2\omega_0$ (red) they coalesce at an exceptional point. For larger damping, the poles move apart along the imaginary axis. Additionally, there are the Matsubara poles (blue) that gain relevance outside the rotating-wave limit, see also the main text. The blue dashed lines correspond to the integration contour used in Eq.~\eqref{eq:G1_under} in the underdamped regime. The arc has an infinite radius.
}\label{fig:poles}
\end{figure}
In the remainder of this section, we want to evaluate the expression given in Eq.~(\ref{eq:G1}).
To fix ideas, we keep $\tau > 0$ since the correlator obeys $G^{(1)}(-\tau)= G^{(1)}(\tau)^*$, \emph{i.e.}, the real part is symmetric while the imaginary part is anti-symmetric in time.
Then, due to causality, the poles with a negative imaginary part that are closest to the real axis are most relevant as the contributions decay the slowest.
The poles are depicted in Fig.~\ref{fig:poles}.

First, there are poles at $\omega_\pm$ with $\omega_\pm = -i\gamma/2 \pm \sqrt{\omega_0^2-\gamma^2/4}$, originating from the poles of $x_\omega$ that dictate the time evolution of $x(t)$, see Eq.~(\ref{eq:x_qle}). 
For weak damping, $\gamma \ll \omega_0$, we have that $\omega_\pm \approx -i\gamma/2 \pm \omega_0$. 
Thus the solution decays at a rate $\gamma/2$ and oscillate with the resonance frequency $\omega_0$.
If the damping is increased, but still in the underdamped regime where $\gamma < 2\omega_0$, the resonance frequency is shifted to $\tilde \omega = \sqrt{\omega_0^2-\gamma^2/4}$.
In the context of open quantum systems, this is also known as the Lamb shift. 
At the critical point, $\gamma = 2\omega_0$, the poles meet on the imaginary axis and coalesce with $\omega_\pm = -i\omega_0$. 
For critical damping, the time evolution is not exponential anymore and is expected to behave like $\omega_0\tau e^{-\omega_0 \tau}$. 

Above the critical point, $\gamma > 2 \omega_0$, we enter the overdamped regime  with $\omega_\pm = -i\gamma/2 \pm i \sqrt{\gamma^2/4-\omega_0^2} $ on the imaginary axis. 
Here, the poles move apart from each other, $\omega_-$ moves away from the real axis while $\omega_+$ approaches $0$.
For strong damping, $\gamma \gg \omega_0$, the poles behave according to $\omega_-\approx -i\gamma$ and $\omega_+ \approx -i\omega_0^2/\gamma \equiv -i\gamma_c$.
Note that in the overdamped regime, the dynamics can be captured by a Fokker-Planck equation \cite{ankerhold:03}.

Besides these poles, the Bose-Einstein distribution provides infinitely many poles at $-i\nu_n = - 2 \pi i k_B T n/\hbar$ with $n\in \mathbb{N}$, with $\nu_n$ the Matsubara frequencies.
They arise due to correlations in the bath where the correlation time is given by $\hbar/k_B T$ and scales inversely proportional to the temperature of the bath. 
This means that the contribution of the Matsubara poles become more relevant as the temperature is lowered (and the poles move closer to the real axis).

For large temperatures, the poles are away from real axis and their contribution is fast decaying.
Due to this, the rotating-wave limit, in which the conventional Lindblad approach is valid, consists of both narrow linewidth $\gamma \ll \omega_0$ and a separation of correlation times $\hbar\gamma \ll k_B T$.
In this case, the damping $\gamma$ provides the smallest rate.
Below, we will denote the rotating-wave limit with $\gamma \to 0^+$.

We will discuss also the opposite limit where $\gamma$ provides the largest rate. 
In this case, a new scale $\gamma_c = \omega_0^2/\gamma \ll \omega_0$ emerges which simultaneously sets the smallest rate, see above.
Thus, in this limit, which we call ultra-strong damping limit ($\gamma \to \infty$), we have $\gamma \gg \omega_0, k_B T/\hbar$ and $\gamma_c \ll k_BT/\hbar$.

Moreover, we will discuss the classical and quantum limit which are defined by the temperature.
In the classical, high-temperature limit, the temperature has to provide the largest rate.
Thus, we require that $k_B T/\hbar \gg \omega, \gamma, \gamma_c$ (we denote this by $T \to \infty$).
The quantum, low-temperature limit describes the situation where the temperature is the smallest rate with $k_B T/\hbar \ll \omega_0, \gamma, \gamma_c$ (we denote this by $T \to 0^+$).

\subsection{Photon current} 
\begin{figure*}
\centering
\includegraphics{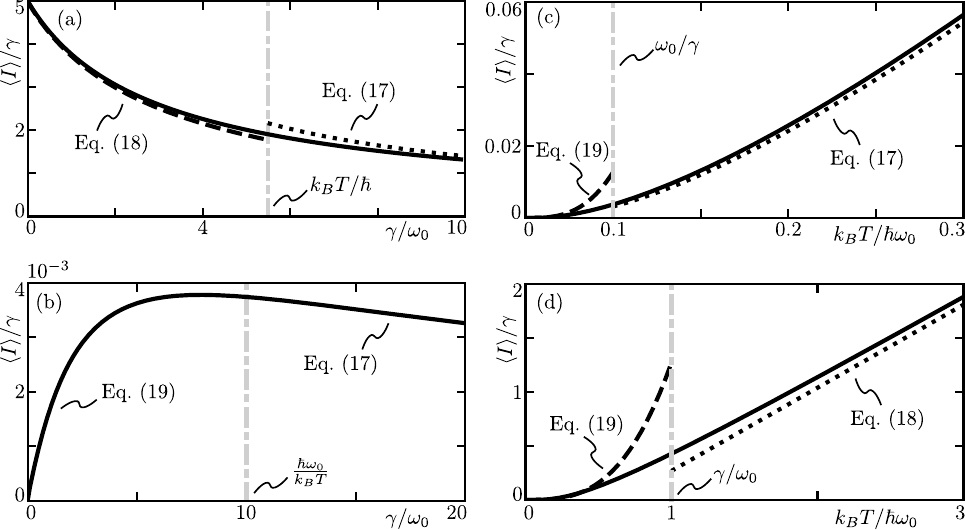}
\caption{
Average photon current $\langle I \rangle$ as a function of damping rate $\gamma$ in (a) and (b), and as a function of temperature $T$ in (c) and (d). 
The full lines display the exact result while dashed and dotted lines represent the approximate results Eq.~\eqref{eq:n_overdamped}-\eqref{eq:low-T} as indicated in the figure. 
(a) Average photon current in the high temperature regime with $k_B T \approx 5.5 \hbar\omega_0$ such that $n_0 = 5$.
(b) Low-temperature regime with $k_BT=0.1\hbar\omega_0$ ($n_0 \approx 5\times 10^{-5}$).
While the photon current is exponentially small in the rotating-wave limit, it increases for stronger damping, Eq.~(\ref{eq:low-T}), until the ultra-strong damping regime of Eq.~(\ref{eq:n_overdamped}) is reached. For both, the high- and low-temperature regimes, the transition to the case of ultra-strong damping is indicated by the gray line.
(c) Average photon current for strong damping with $\gamma=10\omega_0$.
(d) Underdamped regime with $\gamma=\omega_0$.
Going from high $T$ to small $T$, there is a transition to the low-temperature regime, indicated by the gray line.
For (c), there is another crossover from the ultra-strong damping regime~\eqref{eq:n_overdamped} to the high-temperature regime~\eqref{eq:high-T} at $k_BT \approx \hbar\gamma$ (not shown). 
}\label{fig:number}
\end{figure*}

We can obtain the average photon current $\langle I \rangle = \gamma G^{(1)}(0)$, which is also depicted in Fig.~\ref{fig:number}, from Eq.~(\ref{eq:G1}) with
\begin{equation}\label{eq:I}
\langle I \rangle = \int_0^\infty \frac{d\omega}{2\pi} \frac{4 \gamma^2 \omega^2 n_\omega}{(\omega^2-\omega_0^2)^2+\gamma^2\omega^2}.
\end{equation}
In the rotating-wave limit, we have that $\omega_+ \approx -i\gamma/2 + \omega_0$ and $\hbar \gamma \ll k_BT$.
Then, the integrand in (\ref{eq:I}) is strongly localized at $\omega = \omega_0$ with a Lorentzian lineshape of width $\gamma/2$.
This also means that the Bose-Einstein distribution can be evaluated at the resonance frequency $n_{\omega_0}=n_0$.
The average photon current results in ($\gamma \to 0^+$)
\begin{equation}\label{eq:I_rwa}
\langle I \rangle \approx  \gamma n_0 \int_{-\infty}^\infty \frac{d\omega}{2\pi} \frac{\gamma}{(\omega-\omega_0)^2 + \gamma^2/4} = \gamma n_0.
\end{equation}
This reproduces the well-known result from the Lindblad equation where the average photon current is dictated by the temperature of the environment through the Bose-Einstein distribution $n_0$. 

By subtracting the negative frequency tail of the Lorentzian, we can find the first-order correction in $\gamma \ll \omega_0$ to be $\langle I \rangle \approx \gamma n_0(1-\gamma/\pi\omega_0)$.
For an increasing damping rate $\gamma$, the average photon current is suppressed compared to the Lindblad result~(\ref{eq:I_rwa}).
This suppression arises because the environment into which the photons are emitted also effectively acts as a detector.
Larger $\gamma$ mean a larger emission rate and thus also a larger detection rate.
The resulting information gain about the system reduces the uncertainty in the position $x$, which in turn lowers the average photon population inside the system.
Consequently, fewer photons are emitted.

In the ultra-strong damping limit where $\gamma$ provides the largest and $\gamma_c$ the smallest scale, the denominator in (\ref{eq:I}) is dominated by $\gamma^2\omega^2$.
In this limit, the photon current results in ($\gamma \to \infty$)
\begin{equation}\label{eq:n_overdamped}
\langle I \rangle \approx \frac{2}{\pi} \int_{\gamma_c}^\infty d\omega \, n_\omega \approx \frac{2 k_B T}{\pi \hbar} \operatorname{ln}\Bigl(\frac{\gamma k_B T}{\hbar \omega_0^2}\Bigr) + \frac{\omega_0^2}{\pi \gamma},
\end{equation}
Note that the photon current increases logarithmically with the emission rate $\gamma$, while in the rotating-wave limit, see Eq.~(\ref{eq:I_rwa}), the increase is linear.

Next, we study the classical high-temperature regime $k_B T/\hbar \gg \omega_0, \gamma$.
From the Jeans-Rayleigh law for black-body radiation, it is expected that the average photon current is given by $\langle I \rangle/\gamma = k_B T/\hbar\omega_0$.
In this limit, we can expand the Bose-Einstein distribution as $n_\omega\approx k_B T/\hbar\omega -1/2$ where we also keep the first subleading contribution.
We find that the average photon current is given by ($T \to \infty$)
\begin{equation}\label{eq:high-T}
\langle I \rangle \approx \frac{2k_B T\gamma}{\pi\hbar\tilde\omega}\arctan\Bigl(\frac{2\tilde\omega}{\gamma}\Bigr) - \frac\gamma2
\end{equation}
where $\tilde\omega=\sqrt{\omega_0^2-\gamma^2/4}$.
Thus, the equipartition theorem only holds in the rotating-wave limit and is broken outside of this limit.
The expression in (\ref{eq:high-T}) is also valid at the critical point $\gamma=2\omega_0$, where $\tilde\omega=0$, and in the overdamped regime where $\tilde\omega = i\sqrt{\gamma^2/4-\omega_0^2}$, by employing the relations $\lim_{x\rightarrow 0}\arctan(x)/x=1$ and $\arctan(ix)/ix = \operatorname{artanh}(x)/x$ respectively.

At low-temperatures, $k_B T/\hbar \ll \omega_0, \gamma, \gamma_c$, the Bose-Einstein distribution exponentially suppresses excitations at large frequencies, $n_\omega \approx \exp(-\hbar\omega/k_B T)$.
This means that mostly low frequency photons are emitted.
Thus, the integrand in (\ref{eq:I}) can be expanded at $\omega=0$ and we obtain ($T \to 0^+$)
\begin{equation}\label{eq:low-T}
\langle I \rangle \approx \frac{2 \gamma^2}{\pi\omega_0^4} \int_0^\infty d\omega \, \omega^2e^{-\hbar \omega/k_BT} = \frac{4 \gamma^2 (k_B T)^3}{\pi\hbar^3\omega_0^4}.
\end{equation}
While in the high-temperature and the ultra-strong damping limit, the photon current increases slower compared to the rotating-wave limit, in the low-temperature limit the photon current increases quadratically with the damping rate, $\langle I \rangle \propto \gamma^2$.
Note that in both the high- and low-temperature limit, if the damping becomes too large, we have a crossover to the ultra-strong damping limit, see Figs.~\ref{fig:number}(a) and (b). 
Similarly, when the temperature is sufficiently small ($k_BT/\hbar \ll \omega_0,\gamma,\gamma_c$), there is a crossover to the low-temperature regime, see Figs.~\ref{fig:number}(c) and (d).

We see that the average photon current is an analytic  function of $\gamma$, in particular also at the critical point $\gamma=2\omega_0$ where two poles coalesce in an exceptional point. Because of this, in order, to see a signature of  the transition from the under- to the overdamped regime through the point of critical damping, one needs to investigate time-dependent quantities, such as $G^{(1)}(\tau)$.

\subsection{First-order coherence}

As above, we are  first discussing  the rotating-wave limit $\gamma \ll \omega_0, k_BT/\hbar$, where only  a single pole $\omega_+$ is relevant. 
Due to the Lorentzian lineshape, this leads to ($\gamma \to 0^+$)
\begin{equation}\label{eq:G1_RWA}
G^{(1)}(\tau) \approx n_0 e^{-\gamma \tau/2}e^{-i\omega_0 \tau},
\end{equation}
reproducing the well-known result from the Lindblad equation, see  Ref.~\cite{flindt:19}.

Beyond the rotating-wave limit, the contributions of the other poles start to become more important, which leads to a non-trivial time-dependency of $G^{(1)}(\tau)$.
For larger damping, but still in the underdamped regime $\gamma < 2\omega_0$, we can extract the (oscillatory) contribution of $\omega_+$ by a complex contour integration.
We obtain
\begin{multline}\label{eq:G1_under}
G^{(1)}(\tau) = \frac{\omega_+}{\tilde\omega}n_{\omega_+} e^{-\gamma \tau/2}e^{-i\tilde\omega \tau} + \\ \int_0^\infty \frac{d\omega}{2\pi}\frac{4 i \gamma \omega^2 n_{-i \omega}}{(\omega^2+\omega_+^2)(\omega^2+\omega_-^2)}e^{-\omega \tau}.
\end{multline}
The remaining integral arises due to the closure of the contour at the imaginary axis, see Fig.~\ref{fig:poles}.
This contribution leads to a non-exponential time evolution, decaying algebraically for long times, and becomes relevant for larger damping or smaller temperature.

As pointed out above, in the case of ultra-large damping, the new small scale $\gamma_c = \omega_0^2/\gamma \ll \omega_0, k_BT/\hbar$ emerges.
In this limit, the pole $\omega_+ \approx -i\gamma_c$ is the most important as it is closest to the real axis.
In particular, because $\gamma_c \to 0^+$ a separation of time scales is possible, see also Ref.~\cite{hassler:23}.
Reducing the dynamics to $\gamma_c$ leads to ($\gamma \to \infty$)
\begin{equation}
G^{(1)}(\tau) \approx -\frac{k_B T}{2\pi \hbar \gamma}\bigl[e^{x} \operatorname{Ei}(-x) +  e^{-x} \operatorname{Ei}(x) + i\pi e^{-x} \bigr], 
\end{equation}
with $\operatorname{Ei}(x) = \mathcal{P} \int_{-\infty}^x \! dt \, e^{t}/t $ the exponential integral function and $x=\gamma_c\tau$.

Next, we are going to study the case of critical damping ($\gamma = 2\omega_0$).
Classically, it is then expected that $G^{(1)}(\tau)$ assumes a non-exponential time-evolution in the form of $\omega_0\tau e^{-\omega_0 \tau}$.
However, in general, this property is overshadowed by the contributions of the Matsubara poles.
To observe critical damping, one has to go to high temperatures $k_B T\gg \hbar \omega_0$ such that contribution due to the Matsubara frequencies is irrelevant.
In this limit, we obtain ($\gamma=2\omega_0$, $k_BT\gg\hbar\omega_0$)
\begin{multline}
G^{(1)}(\tau) \approx \frac{k_B T}{\pi \hbar \omega_0}\bigl[ 2 + y e^{y} \operatorname{Ei}(-y) - y e^{-y} \operatorname{Ei}(y)\bigr] \\
-\frac{i k_B T}{\hbar \omega_0} y e^{-y}
\end{multline}
with $y=\omega_0\tau$.
Note that in the classical limit, the critical damping behavior can be readily observed in the imaginary part of $G^{(1)}(\tau)$.

As a last limit, we investigate the low-temperature regime $k_BT/\hbar \ll \omega_0, \gamma, \gamma_c$.
Here, the contributions of the Matsubara poles dominate which leads to ($T \to 0^+$)
\begin{equation}\label{eq:G1_lowT}
G^{(1)}(\tau) \approx \frac{4 \gamma (k_B T)^3}{\pi\hbar^3\omega_0^4} \frac{1}{(1+i\hbar \tau/k_B T)^3}.
\end{equation}
We find that the correlations that are induced by the bath have a purely algebraic decay.
For long times, $\tau \gg \hbar/k_B T$, the coherence function will thus decay according to $\operatorname{Im}[G^{(1)}(\tau)] \propto \tau^{-3}$ and $\operatorname{Re}[G^{(1)}(\tau)] \propto \tau^{-4}$.

\section{Effective Lindblad equation}\label{sec:lindblad}

\begin{figure}[tb]
	\centering
	\includegraphics{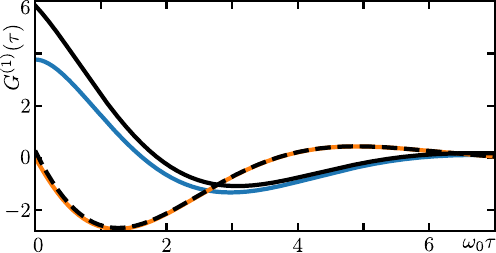}
	\caption{%
	Comparison of the exact solution of $G^{(1)}(\tau)$ from Eq.~(\ref{eq:G1}) (real part in blue, imaginary part in orange), to the approximate solution given by the effective Lindblad equation~(\ref{eq:Lindblad_eff}), in black (real part as full line, imaginary part dashed), outside the rotating-wave limit with $\gamma = \omega_0$ and $k_BT \approx 5.5\hbar \omega_0$ ($n_0=5$).
Note that the imaginary part is well approximated for all times, while the real part provides a good approximation for $\gamma \tau \gtrsim 1$, as expected. 
}\label{fig:g1}
\end{figure}
As mentioned throughout this work, the Lindblad equation relies on a RWA which only works for weak damping $\gamma \ll \omega_0, k_B T/\hbar$.
In this limit, the dynamics can be solely described by the pole $\omega_+$ leading to a Lindblad equation.
For larger damping, but in the underdamped regime $\gamma < 2\omega_0$, there is still a contribution due to $\omega_+$, see first term in Eq.~(\ref{eq:G1_under}).
In order to capture this contribution, we can perform a RWA with respect to the Lamb-shifted frequency  $\tilde \omega = \sqrt{\omega_0^2-\gamma^2/4}$. 
Because of this, we are able to retrieve an effective Lindblad equation outside of the rotating-wave limit which  captures the coherent dynamics, similar to the procedure introduced in Ref.~\cite{hassler:19}.

To this end, we define a set of new ladder operators $\tilde{a}$ and $\tilde a ^\dag$ that are associated to $\omega_+$ at the frequency $\tilde\omega$. 
Similar to the RWA, they are defined by  $a = e^{-i\tilde\omega t}\tilde a$ and obey the commutation relation $[\tilde a, \tilde a^\dag] = 1$. The photon current is then given by $I = r\gamma\,  \tilde a^\dag \tilde a$ with $r= \omega_+ n_{\omega_+}/\tilde \omega n_{\tilde\omega}$.

Before we move on, we explain the meaning of $r$.
It is akin to the quasiparticle weight, well-known in the field of many-body physics \cite{orland}.
The quasiparticle weight provides a measure for the contribution of the pole $\omega_+$ and thus also quantifies the validity of the resulting Lindblad equation.
In the rotating-wave limit ($\gamma \to 0^+$), we have that $r=1$ as the dynamics are accurately described by a Lindblad equation.
Moving beyond this limit, the quasiparticle weight reads ($\hbar \gamma \ll \hbar \omega_0 \ll k_B T$) \cite{note_quasi}
\begin{equation}
r = \frac{\omega_+ n_{\omega_+}}{\tilde \omega n_{\tilde\omega}} \approx 1 + i\frac{\hbar\gamma}{4k_BT} - \frac{1}{48}\Bigl(\frac{\hbar \gamma}{k_BT}\Bigr)^2.
\end{equation}
The deviation from $1$ for increasing damping or lower temperatures is due to the fact that the importance of the Matsubara poles increases.

With the procedure explained above, the time evolution of the density matrix of the system $\rho$ is dictated by the effective Lindblad equation
\begin{equation}\label{eq:Lindblad_eff}
\dot \rho =  \gamma (n_{\tilde\omega}+1)\mathcal{D}[\tilde a]\rho +  \gamma n_{\tilde\omega} \mathcal{D}[\tilde a^\dag]\rho,
\end{equation}
where $\mathcal{D}[L]\rho = L\rho L^\dag - \tfrac12\{L^\dag L, \rho\}$.
Note that the (complex) quasiparticle weight does not enter the effective Lindblad equation.
Thus, the time evolution remains completely positive and trace preserving.

We expect the effective Lindblad equation to be valid on an intermediate timescale, akin to Fermis' golden rule. 
For short times, all poles are important and the effective Lindblad equation is not able to capture the dynamics accurately as it only takes a single pole into account.
For long times, the coherence function decays algebraically, while the Lindblad equation reproduces an exponential decay.
However, as the coherence function is already small for such long times, the Lindblad equation provides the relevant contribution for $\gamma\tau \gtrsim 1$.

The resulting $G^{(1)}(\tau)$, the first part of Eq.~(\ref{eq:G1_under}), is depicted in Fig.~\ref{fig:g1} and is compared to the exact result obtained from Eq.~(\ref{eq:G1}).
We see that the imaginary part of $G^{(1)}(\tau)$ is well approximated for all times, while the real part provides a good approximation for $\gamma\tau \gtrsim 1$.

\section{Conclusion}
In conclusion, we have shown how to obtain the radiation statistics of an arbitrary system beyond the rotating-wave approximation, where the  Lindblad master equation is not applicable.
For bosonic systems, we have pointed out that emission of photons cannot be described by the usual definition of a ladder operator via position and momentum as this leads to a diverging number of `photons'.
Instead, the dissipative processes have to be characterized by the frequency response, emission of photons at positive frequencies and absorption at negative frequencies.
Importantly, we have provided a general expression for the photon current operator for arbitrary systems and environments which allows to study its radiation statistics.
As a demonstration of our formalism, we have calculated the radiation statistics of a damped harmonic oscillator coupled to a thermal environment beyond the RWA. 
We  obtained finite results for both the photon current and the correlation function.
Moreover, we have shown how to obtain an effective Lindblad master equation that is valid outside the rotating-wave limit. For this, we have introduced a quasiparticle weight, measuring the coherent contribution to the response, similarly to the notion in many-body systems.

This work provides a framework for further research where the method can be applied to a variety of systems and extended in different ways.
For example, one could study driven cavities similar to Ref.~\cite{portugal:23}, the impact of nonlinear potentials \cite{khmochenko}, or the radiation statistics of two-level systems \cite{leggett:87}.
Further, one could extend our formalism for the study of quantum thermodynamics beyond the RWA   \cite{potts}.

\acknowledgements

This work was supported by the Deutsche Forschungsgemeinschaft (DFG) under Grant No. HA 7084/8–1.

\appendix

\section{Derivation of photon current operator}
In this appendix, we show how to obtain the photon current operator~\eqref{eq:photon_current} starting from the dissipated power~\eqref{eq:power}.
From Eq.~\eqref{eq:power}, the power can also be written as 
\begin{equation}
P(t) = -\frac12 \int\!\!\!\!\int \frac{d\omega d\omega'}{(2\pi)^2} \omega \omega' f_{\omega'} \{ X_\omega, X_{\omega'} \} e^{-i(\omega+\omega')t}.
\end{equation}
Note that we can simplify this expression by symmetrizing $f_\omega$ with respect to $\omega$ and $\omega'$ by relabeling the frequencies. 
Then, the expression will be symmetric for $\omega \leftrightarrow \omega'$ which allows us to write $\{X_\omega, X_{\omega'}\} = 2X_\omega X_{\omega'}$ in the integral.
This leads to 
\begin{equation}
P(t) = -\frac12\int\!\!\!\!\int \frac{d\omega d\omega'}{(2\pi)^2} \omega\omega'(f_\omega+f_{\omega'})X_\omega X_{\omega'} e^{-i(\omega+\omega')t}.
\end{equation}
Now, we redefine $\omega \rightarrow -\omega$ and introduce the average frequency $\Omega=(\omega+\omega')/2$ and the frequency difference $\nu = \omega-\omega'$.
We obtain
\begin{multline}
P(t) = \frac12\int\!\!\!\!\int \frac{d\Omega d\nu}{(2\pi)^2} \sqrt{\left(\Omega+\frac\nu2\right)\left(\Omega-\frac\nu2\right)} \\ \times (f^*_{\Omega+\frac\nu2}+f_{\Omega-\frac\nu2}) A^\dag_{\Omega+\frac\nu2} A_{\Omega-\frac\nu2} e^{i\nu t},
\end{multline}
where we employed the relations $f_{-\omega}=f^*_{\omega}$ and $X_{-\omega}=X^\dag_{\omega}$. 
Furthermore, we introduced the emission operators $A_\omega = \Theta(\omega) \sqrt{|\omega|} X_\omega$ with $A_{-\omega} = A_\omega^\dagger$ as described in the main text.
The restriction to positive frequencies ensures to describe only emission events~\cite{glauber, clerk, lesovik:97}.

At this point, the expression is still exact and describes that energy is mostly emitted at a frequency $\Omega$ with fluctuations $\nu$.
To resolve the emission at the frequency $\Omega$, the detection time is required to be large such that $\Omega t \gg 1$.
Then, only small $\nu$ will be relevant, which we keep in the operators to capture the fluctuating dynamics of the system.
We obtain
\begin{equation}
P(t) \approx \int\!\!\!\!\int \frac{d\Omega d\nu}{(2\pi)^2} \Omega \operatorname{Re}(f_\Omega) A^\dag_{\Omega+\frac\nu2} A_{\Omega-\frac\nu2} e^{i\nu t},
\end{equation}
which describes the energy current.
After dividing the integrand by $\hbar \Omega$, we obtain the photon current operator $I(t)$ as given in the main text, see Eq.~\eqref{eq:photon_current}.


\begin{thebibliography}{10}
\bibitem{note_damping}
We include the infinitesimal time delay by $0^+$ for causality.
\bibitem{note_quasi}
In contrast to the many-body setting, $r$ is complex which is due to the non-Hermitian nature of the problem.
\end{thebibliography}
\end{document}